\documentstyle[12pt,a4wide]{article}
\begin{document}
\author{B. Boisseau\thanks{E-mail : 
boisseau@celfi.phys.univ-tours.fr}, and B. Linet\thanks{E-mail : 
linet@celfi.phys.univ-tours.fr} \\
\small Laboratoire de Math\'ematiques et Physique Th\'eorique \\
\small CNRS/UPRES-A 6083, Universit\'e Fran\c{c}ois Rabelais \\
\small Facult\'e des Sciences et Techniques \\
\small Parc de Grandmont 37200 TOURS, France}
\title{\bf Exact metric for the exterior of a global string 
in the Brans-Dicke theory}
\date{}
\maketitle

\begin{abstract}

We determine in closed form the general static solution with cylindrical
symmetry to the Brans-Dicke equations for an energy-momentum tensor
corresponding to the one of the straight U(1) global string outside the
core radius assuming that the Goldstone boson field takes its
asymptotic value.

\end{abstract}

\section{Introduction}

Topological defects could be produced at a phase transition in the early 
universe \cite{kib,vil}. Their nature depends on the symmetry
broken in the field theory under consideration. A class of topological 
defects are the global defects as the global strings which are not
finite energy. So, a static, straight U(1) global string in general relativity
has a metric which is not asymptotically 
Minkowskian \cite{coh,gre1,har1,gib1}. Indeed, the spacetime
has necessarily a physical singularity at a finite proper distance of the axis 
\cite{gre1} giving constraints on the abundance of the global strings 
in the early universe \cite{lar}. 
The explicit metric outside the core radius within the approximate theory 
in which the Goldstone boson field takes its asymptotic value has been 
obtained by Cohen and Kaplan \cite{coh} and it presents a curvature 
singularity at a finite proper distance. 

It is widely accepted that a gravitational scalar field, beside the metric
of the spacetime, must exist in the framework of the present unified
theories. These scalar-tensor theories of gravitation 
take their importance in the early universe where
it is expected that the coupling to matter of the scalar field would be as 
same order
as the one of metric althought the scalar coupling is negligeable in
the present time. Now topological defects are produced during vacuum phase
transitions in the early universe, therefore several authors studied the
static solutions generated for instance by a  straight U(1) gauge string in 
the Brans-Dicke theory \cite{gun,bar}, in the scalar-tensor theories 
with matter minimally coupled \cite{gui} or 
in the dilaton theories \cite{gre2}. Of course the scalar field is supposed
massless because in the massive case the theory is pratically general
relativity for distance much larger than the range of the scalar field.

Recently, Sen {\em et al} \cite{sen} studied the static, straight 
U(1) global string in the
Brans-Dicke theory but they did not arrive to determine in closed form
the general solution outside the core radius of the straight U(1) global
string.
Also, we take up again the problem of the determination of the exact solution
outside the core radius in the case where the Goldstone boson field takes
its asymptotic value.

The gravitational field variables of the Brans-Dicke theory are the metric 
$\hat{g}_{\mu \nu}$ of the spacetime and a scalar field $\hat{\phi}$. 
The matter is minimally coupled to $\hat{g}_{\mu \nu}$ and its
energy-momentum tensor $\hat{T}^{\mu}_{\nu}$ is conserved. 
Nevertheless, it is
now well-known that it is more convenient to use a non-physical metric
$g_{\mu \nu}$ and a new scalar $\phi$ defined by
\begin{equation}
\label{1}
\hat{g}_{\mu \nu}=\exp (2\alpha \phi )g_{\mu \nu} \quad 
{\rm and}\quad \hat{\phi}=\frac{1}{\cal G}\exp (-2\alpha \phi )
\end{equation}
where the gravitational coupling constant ${\cal G}$ can be related to the
Newtonian constant. The parameter $\alpha$ in (\ref{1}) has the value
\[
\alpha{^2}=\frac{1}{2\omega +3}
\]
in terms of the usual Brans-Dicke parameter $\omega$. We also introduce
a non-physical source $T_{\mu \nu}$ defined by
\begin{equation}
\label{2}
T_{\mu \nu}=\exp (2\alpha \phi )\hat{T}_{\mu \nu}.
\end{equation}

We describe the straight U(1) global string by a static, cylindrically 
symmetric spacetime. Consequently, we can write the non-physical metric
in the form
\begin{equation}
\label{3}
ds^2=d\rho^2+g_2(\rho )dz^2+g_3(\rho )d\varphi^2-g_4(\rho )dt^2 
\end{equation}
in the coordinate system $(\rho ,z,\varphi ,t)$ with $\rho >0$ and
$0\leq \varphi <2\pi$ where the functions $g_2$, $g_3$ and $g_4$ are strictly 
positive. The scalar field depends only on $\rho$. Under our assumptions, 
the form of the energy-momentum tensor $\hat{T}^{\mu}_{\nu}$ outside the 
core radius of the U(1) global string yields 
\begin{equation}
\label{4}
T^{\rho}_{\rho}=T^{z}_{z}=T^{t}_{t}=-T^{\varphi}_{\varphi}=-\sigma (\rho )
\end{equation}
where the strictly positive function $\sigma$ is to be determined for 
$\rho >\rho_C$, $\rho_C$ being the core radius. As noticed by 
Gibbons {\em et al} \cite{gib2}, form (\ref{4}) corresponds also 
to a $\sigma$-model with vanishing potential
and having a target space with closed geodesics.

The purpose of this work is to give the general expression of static metrics 
with cylindrical symmetry
and the scalar field which are the solutions to the Brans-Dicke equations  
with a source having the algebraic form (\ref{4}). In the particular
case of a straight U(1) global string, metric (\ref{3}) must exhibit the
boost invariance, i.e. $g_2=g_4$.

The plan of the work is as follows. In Section 2, we give the basic
equations of our problem which are to be solved. The explicit solutions
are obtained in Section 3. We discuss in Section 4 the singularities
of the solutions and the existence of black hole
solutions. We add in Section 5 some concluding remarks.

\section{Gravitational field equations}

In terms of the non-physical metric  $g_{\mu \nu}$   
and the scalar field $\phi$ introduced by relations (\ref{1}), 
the Brans-Dicke equations are
\begin{equation}
\label{5}
R_{\mu \nu}=2\partial_{\mu}\phi \partial_{\nu}\phi +8\pi {\cal G}
(T_{\mu \nu}-\frac{1}{2}Tg_{\mu \nu}),
\end{equation}
\begin{equation}
\label{6}
\Box \phi =-4\pi {\cal G}\alpha T
\end{equation}
where the source is the non-physical energy-momentum tensor (\ref{2}).
We derive from (\ref{5}) and (\ref{6}) that
\begin{equation}
\label{7}
\nabla_{\mu}T^{\mu}_{\nu}=\alpha T\partial_{\nu}\phi .
\end{equation}
We note that equation (\ref{5}) can be rewritten
\begin{equation}
\label{8}
G_{\mu \nu}=R_{\mu \nu}-\frac{1}{2}Rg_{\mu \nu}=2\partial_{\mu}\phi 
\partial_{\nu}\phi
-g_{\mu \nu}g^{\gamma \delta}\partial_{\gamma}\phi \partial_{\delta}\phi
+8\pi {\cal G}T_{\mu \nu}.
\end{equation}

For metrics (\ref{3}) with source (\ref{4}), equation (\ref{7}) reduces to
\[
\frac{d\sigma}{d\rho} +\frac{\sigma}{g_3}\frac{dg_3}{d\rho} =
2\alpha \sigma \frac{d\phi}{d\rho}
\]
whose the general solution has the form
\begin{equation}
\label{9}
\sigma (\rho )=\frac{\sigma_0}{8\pi {\cal G}}
\frac{\exp (2\alpha \phi (\rho ))}{g_3(\rho )}
\end{equation}
where $\sigma_0$ is an arbitrary positive constant.

We are now in a position to write down the gravitational field equations. 
In equation (\ref{5}), the 
components $R^{z}_{z}$, $R^{\varphi}_{\varphi}$ and $R^{t}_{t}$ of the Ricci
tensor give firstly
\begin{equation}
\label{10}
\frac{d}{d\rho}\left( \frac{u}{g_2}\frac{dg_2}{d\rho}\right) =0,
\end{equation}
\begin{equation}
\label{11}
\frac{d}{d\rho}\left( \frac{u}{g_3}\frac{dg_3}{d\rho} \right) =-4\sigma_0
\frac{u\exp (2\alpha \phi )}{g_3},
\end{equation}
\begin{equation}
\label{12}
\frac{d}{d\rho}\left( \frac{u}{g_4}\frac{dg_4}{d\rho} \right) =0
\end{equation}
where $u$ denotes the square root of the determinant, $u^2=g_2g_3g_4$. 
Secondly, the scalar equation (\ref{6}) is written as
\begin{equation}
\label{13}
\frac{d}{d\rho}\left( u\frac{d\phi}{d\rho}\right) =\alpha\sigma_0
\frac{u\exp (2\alpha \phi )}{g_3}.
\end{equation}
In equation (\ref{8}), the component $G^{\rho}_{\rho}$ of the Einstein 
tensor gives a constraint equation
\begin{equation}
\label{14}
\frac{1}{g_2g_3}\frac{dg_2}{d\rho}\frac{dg_3}{d\rho}+\frac{1}{g_3g_4}
\frac{dg_3}{d\rho}\frac{dg_4}{d\rho}+\frac{1}{g_4g_2}\frac{dg_4}{d\rho}
\frac{dg_2}{d\rho}=-4\sigma_0 \frac{\exp (2\alpha \phi )}{g_3}+
4\left( \frac{d\phi}{d\rho} \right)^{2}.
\end{equation}
We have a system of five differential equations for $g_2$, $g_3$, $g_4$ 
and $\phi$ which are compatible since we have taken a $T_{\mu \nu}$
satisfying identically the integrability condition (\ref{7}).

In order to solve these equations we introduce a new radial coordinate 
$r$ related to $\rho$ by
\begin{equation}
\label{15}
u(\rho )\frac{dr}{d\rho}=1.
\end{equation}
We have $-\infty <r<\infty$ when $0<\rho <\infty$ since {\em a priori} the
integral of $1/u$ diverges as $\rho \rightarrow 0$. In this
coordinate system  metric (\ref{3}) has the form
\begin{equation}
\label{3a}
ds^2=g_2(r)g_3(r)g_4(r)dr^2+g_2(r)dz^2+g_3(r)d\varphi^2 -g_4(r)dt^2.
\end{equation}
Of course, the energy-momentum tensor keeps form (\ref{4}).
The gravitational field equations (\ref{10}-\ref{14}) become
\begin{equation}
\label{16}
\frac{d}{dr}\left( \frac{1}{g_2}\frac{dg_2}{dr}\right) =0,
\end{equation}
\begin{equation}
\label{17}
\frac{d}{dr}\left( \frac{1}{g_3}\frac{dg_3}{dr} \right) =-4\sigma_0
g_2g_4\exp (2\alpha \phi ),
\end{equation}
\begin{equation}
\label{18}
\frac{d}{dr}\left( \frac{1}{g_4}\frac{dg_4}{dr} \right) =0,
\end{equation}
\begin{equation}
\label{19}
\frac{d^2\phi}{dr^2}=\alpha \sigma_0g_2g_4\exp (2\alpha \phi ),
\end{equation}
\begin{equation}
\label{20}
\frac{1}{g_2g_3}\frac{dg_2}{dr}\frac{dg_3}{dr}+\frac{1}{g_3g_4}
\frac{dg_3}{dr}\frac{dg_4}{dr}+\frac{1}{g_4g_2}\frac{dg_4}{dr}
\frac{dg_2}{dr}=-4\sigma_0 g_2g_4\exp (2\alpha \phi )+4
\left( \frac{d\phi}{dr} \right)^{2}.
\end{equation}

\section{Explicit determination of the solutions}

Equations (\ref{16}) and (\ref{18}) have the obvious solutions
\begin{equation}
\label{21}
g_2(r)=g_2^0\exp (K_2r) \quad {\rm and}\quad g_4(r)=g_4^0\exp (K_4r)
\end{equation}
where $g_2^0$, $K_2$, $g_4^0$ and $K_4$ are arbitrary constants.  
By combining equations (\ref{17}) and (\ref{19}) we obtain the relation
\begin{equation}
\label{22}
g_3(r)=g_3^0 \exp (K_3r)\exp\left( -\frac{4}{\alpha}\phi \right)
\end{equation}
where $g_3^0$ and $K_3$ are arbitrary constants. Hereafter we denote $g_3$ 
by $g$. We have now to distinguish two cases.

\subsection{Case $K_2=0$ and $K_4=0$}

In the case where $K_2=0$ and $K_4=0$, metric (\ref{3a}) reduces to
\begin{equation}
\label{3b}
ds^2=g(r)dr^2+dz^2+g(r)d\varphi^2-dt^2
\end{equation}
in rescaled coordinates $r$, $z$ and $t$.
It remains three equations of system (\ref{16}-\ref{20}). Equation (\ref{20})
leads to
\begin{equation}
\label{20b}
\frac{d\phi}{dr}=\pm \sqrt{\sigma_0}\exp (\alpha \phi )
\end{equation}
and from this equation (\ref{19}) is automatically satisfied. We can integrate
equation (\ref{20b}) and we get the expression of the scalar field
\begin{equation}
\label{20c}
\exp (2\alpha \phi )=\frac{1}{\alpha^2\sigma_0(r-k)^2}
\end{equation}
where $k$ is an arbitrary constant. Taking into account (\ref{22}) we now 
obtain the components of metric (\ref{3b})
\begin{equation}
\label{22b}
g(r)=g^0\exp (Kr)\mid r-k\mid ^{4/\alpha^2}.
\end{equation}

\subsection{Case $K_2\not= 0$ or $K_4\not= 0$}

When $K_2\not= 0$ or $K_4\not=0$,
it is always possible to write $K_2=\mu (1-w)$ and $K_4=\mu (1+w)$ for
specific constants $\mu$ and $w$, $\mu \not= 0$. We then 
introduce a dimensionless coordinate $x$ by setting
\begin{equation}
\label{23}
x=\exp (\mu r)   
\end{equation}
where $0<x<\infty$. If $\mu >0$ then the coordinate $x$ is increasing with $r$.
In this coordinate system metric (\ref{3}) has the form
\[
ds^2=\frac{1}{\mu^2}g_2^0g_4^0g_3(x)dx^2+g_2^0x^{1-w}dz^2+g_3(x)d\varphi^2-
g_4^0x^{1+w}dt^2.
\]
By a rescaling of the coordinates $x$, $z$ and $t$, we can put this metric 
in the following form 
\begin{equation}
\label{25}
ds^2=g(x)dx^2+x^{1-w}dz^2+g(x)d\varphi^2-x^{1+w}dt^2.
\end{equation}
The energy-momentum tensor keeps form (\ref{4}) and $\sigma$ is given 
by (\ref{9}).

It is convenient to write directly the field equations (\ref{5}) and 
(\ref{6}) for metric (\ref{25}); we thus obtain
\begin{equation}
\label{27}
\frac{1}{x^2}-\frac{w^2}{x^2}+
\frac{g'}{xg}+\frac{(g')^2}{g^2}-\frac{g''}{g}=4(\phi ')^2,
\end{equation}
\begin{equation}
\label{28}
\frac{(g')^2}{g^2}-\frac{g''}{g}-\frac{g'}{xg}=4\sigma_0\exp (2\alpha \phi ),
\end{equation}
\begin{equation}
\label{29}
\frac{1}{x}\left( x\phi '\right) '=\alpha \sigma_0\exp (2\alpha \phi ),
\end{equation}
the other components being identically verified. 
Moreover relation (\ref{22}) is now written as
\begin{equation}
\label{30}
g(x)=g^0 x^c \exp \left( -\frac{4}{\alpha}\phi \right)
\end{equation}
where $c$ are an arbitrary constant.

We now introduce the function $y(x)$ defined by $y(x)=2\alpha \phi$.
Taking into account (\ref{30}), the system of differential equation 
(\ref{27}-\ref{29}) reduces to 
\begin{equation}
\label{32}
2x^2y''-2xy'-x^2y'^2+\alpha^2(1-w^2+2c)=0,
\end{equation}
\begin{equation}
\label{33}
y''+\frac{1}{x}y'=2\alpha^2\sigma_0 \exp y
\end{equation}
for the unknowm function $y$.
In the appendix, we give the explicit expression of the common solutions to
the differential equations (\ref{32}) and (\ref{33}) by setting
$C=\alpha^2(1-w^2+2c)$.

To summarise this, the desired 
metric (\ref{25}), for a given $w$ and $g^0$, and the scalar field 
$\phi$ depends on three constants $C_1$, $C_2$ and $n$ because we express
$c$ in terms of $n$ and $w$.
A convenient classification of the solutions is to use the sign of 
$n$; we obtain thereby for $n>0$
\begin{eqnarray}
\label{40}
& & \nonumber g(x)=g^0 x^{(2/\alpha^2-1/2+w^2/2+2n^2/\alpha^2)}
[\mid Z(x)\mid]^{4/\alpha^2} \\
& & \phi (x) =-\frac{1}{\alpha}\left( \ln x+\ln \mid Z(x)\mid \right) \\
& & \nonumber {\rm with}\quad Z(x)=C_1x^n+C_2x^{-n}\quad C_1C_2<0,
\end{eqnarray}
for $n=0$
\begin{eqnarray}
\label{41}
& & \nonumber g(x)=g^0 x^{(2/\alpha^2-1/2+w^2/2)}
[\mid Z(x)\mid ]^{4/\alpha^2} \\
& & \phi (x) =-\frac{1}{\alpha}\left( \ln x+\ln \mid Z(x)\mid \right) \\
& & \nonumber {\rm with}\quad Z(x)=C_1+C_2\ln x \quad C_2\not= 0
\end{eqnarray}
and for $n<0$
\begin{eqnarray}
\label{42}
& & \nonumber g(x)=g^0 x^{(2/\alpha^2-1/2+w^2/2-2n^2/\alpha^2)}
[\mid Z(x)\mid ]^{4/\alpha^2} \\
& & \phi (x)=-\frac{1}{\alpha}\left( \ln x+\ln \mid Z(x)\mid \right) \\
& & \nonumber {\rm with} \quad Z(x)=C_1\sin (n\ln x)+C_2\cos (n\ln x) \quad
C_1\not= 0 \quad {\rm or}\quad C_2\not= 0.
\end{eqnarray}
The strictly positive value of $\sigma_0$ characterizing the energy-momentum 
tensor (\ref{9}) is given in the appendix by relations (\ref{39}) in terms of
the constants $C_1$, $C_2$ and $n$ appearing in solutions (\ref{40}-\ref{42}).

\section{Singularities of the solutions and black hole cases}

\subsection{Case $K_2=0$ and $K_4=0$}

We write down the physical metric (\ref{1}) associated with metric (\ref{3b})
where the function $g$ is given by (\ref{22b})
\begin{equation}
\label{51}
d\hat{s}^2=\frac{g^0}{\alpha^2\sigma_0}
\mid r-k\mid^{4/\alpha^2-2}\exp (Kr)\left( dr^2+d\varphi^2\right)
+\frac{1}{\alpha^2\sigma_0(r-k)^2}\left( dz^2-dt^2\right)
\end{equation}
with $-\infty <r<\infty$ in principle. It is obvious that the Riemann tensor
of metric (\ref{51}) diverges at $r=k$. So, there exists two intervals of
definition of the metric : $-\infty <r<k$ and $k<r<\infty$. The point $r=k$
is at a finite proper distance in the two domains $r<k$ and $r>k$
since the values of the proper radial coordinate, 
respectively given by the integrals 
\[
\int^{r}(-r+k)^{2/\alpha^2-1}\exp (Kr/2)dr \quad {\rm and}\quad
\int_{r}(r-k)^{2/\alpha^2-1}\exp (Kr/2)dr
\]
are finite as $r \rightarrow k$. We see from (\ref{20c}) that the scalar 
field is also singular at $r=k$.

\subsection{Case $K_2\not= 0$ or $K_4\not= 0$}

In this case the physical metric (\ref{1}) associated with metric (\ref{25})
has the form
\begin{equation}
\label{52}
d\hat{s}^2=\frac{g(x)}{x^2Z^2(x)}\left( dx^2+d\varphi^2\right)
+\frac{x^{-1-w}}{Z^2(x)}dz^2-\frac{x^{-1+w}}{Z^2(x)}dt^2
\end{equation}
where the functions $Z$ and $g$ are given in (\ref{40}-\ref{42}) with in
principle $0<x<\infty$. It is clear from the expression of the components
of metric (\ref{52}) that the zeros of the function $Z$ are the singularities 
of the Riemann tensor. They are at finite proper distances. 
The scalar field is also singular at the zeros of the function $Z$.

The case $x=0$ yields a singularity for metric (\ref{52})  
except if $x=0$ is an horizon. This
eventuality occurs only possible for solution (\ref{40}) when 
$-1+w+2n>0$ since the component $g_{tt}$ is proportional to
$x^{-1+w+2n}$ as $x\rightarrow 0$.
To obtain a black hole, we must have that the components $g_{zz}$ and 
$g_{\varphi \varphi}$ 
remain constant as $x\rightarrow 0$. Therefore we require 
\begin{equation}
\label{bh}
w=2n-1 \quad {\rm and} \quad 
2/\alpha^2-1/2+w^2/2+2n/\alpha^2-2-4n/\alpha^2+2n=0.
\end{equation}
As a consequence, $g_{tt}$ is proportional to $x^{4n-2}$ as 
$x\rightarrow 0$. We now perform the following change of radial coordinate
\[
R(x)=\frac{\sqrt{g(x)}}{xZ(x)}.
\]
Taking into account the previous relations, we find that the components 
$g_{RR}$ is proportional to $x^{-4n+2}$ as $x\rightarrow 0$. Now there
exists certainly a coordinate system in which the components of the metric
are regular at $x=0$, therefore we obtain a family of black hole metrics
(\ref{52}) when the constants verify relations (\ref{bh}). The situation
is similar in general relativity where a black hole metric exists for $w=1$ 
\cite{har2}. However, the scalar field is only regular at $x=0$ for $n=1/2$.
Also, we have not really a solution to Brans-Dicke equations representing
a black hole.

\section{Conclusion}

We have explicitly found the general static solution with cylindrical
symmetry to the Brans-Dicke equations with a source having the  algebraic form
(\ref{4}).
There are two classes of solutions: metrics (\ref{51}) and metrics (\ref{52}).

The general static metric describing a straight global string 
outside the core radius is obtained
by requiring that $g_2=g_4$. We write down the physical metric
(\ref{1}) for the two classes. Firstly, we have directly metric (\ref{51})
\begin{eqnarray}
& & \nonumber d\hat{s}^2=\frac{g^0}{\alpha^2\sigma_0}
 \mid r-k\mid^{4/\alpha^2-2}\exp (Kr)
\left( dr^2+d\varphi^2\right)+\frac{1}{\alpha^2\sigma_0(r-k)^2}
\left( dz^2-dt^2\right) , \\
& & \hat{\phi}=\frac{\alpha^2\sigma_0}{\cal G} (r-k)^2.
\end{eqnarray}
This form of metric does not exist in general relativity.
Secondly, by setting $w=0$ in metric (\ref{52}) we get
\begin{eqnarray}
& & \nonumber d\hat{s}^2=\frac{1}{x^2 Z^2(x)}\left[ g(x)
\left( dx^2+d\varphi^2 \right)+x\left( dz^2-dt^2\right) \right] , \\
& & \hat{\phi}=\frac{1}{\cal G}x^2Z^2(x).
\end{eqnarray}   
where the functions $g$ and $Z$ are given in (\ref{40}-\ref{42}) with $w=0$.
We have proved that the point $x=0$ and the zeros of the function $Z$ are
physical singularities of the solutions in the Brans-Dicke theory. 
We have not touched upon the
question concerning the matching of these solutions with the asymptotic
solutions describing a straight U(1) global string. 

\section*{Appendix}

Following Kamke \cite{kam} p. 562, it is possible to find the common 
solutions to the following system of differential equations
\begin{equation}
\label{32c}
y''+\frac{1}{x}y'=2\alpha^2 \sigma_0\exp y ,
\end{equation}
\begin{equation}
\label{33c}
2x^2y'' -2xy' -x^2 y'^2+C=0
\end{equation}
where $C$ is a constant.
We firstly solve equation (\ref{33c}). By means of the change of function
\begin{equation}
\label{34}
y(x)=-2\ln x -2\ln \mid Z(x)\mid ,
\end{equation}
we derive the Euler equation
\begin{equation}
\label{35}
x^2Z''+xZ'-[1+\frac{1}{4}C]Z=0.
\end{equation}
The expression of the general solution to equation (\ref{35}), valid for 
$x>0$, depends on the sign of $1+C/4$
\begin{enumerate}
\item If $1+C/4>0$ then we get

\begin{equation}
\label{36}
Z^{(1)}(x)=C_1^{(1)} x^n +C_2^{(1)} x^{-n} \quad {\rm with}\quad 
n=\sqrt{1+C/4}
\end{equation}
where $C_1^{(1)}$ et $C_2^{(1)}$ are constants of integration.
\item If $1+C/4=0$ then we get
\begin{equation}
\label{37}
Z^{(2)}(x)=C_1^{(2)}+C_2^{(2)}\ln x 
\end{equation}
where $C_1^{(2)}$ and $C_2^{(2)}$ are constants of integration.
\item If $1+C/4<0$ then we get
\begin{equation}
\label{38}
Z^{(3)}(x)=C_1^{(3)}\sin (n\ln x)+C_2^{(3)}\cos (n\ln x) \quad 
{\rm with} \quad n=-\sqrt{-1-C/4}
\end{equation}
where $C_1^{(3)}$ and $C_2^{(3)}$ are constants of integration.
\end{enumerate}

We now verify that solutions (\ref{36}-\ref{38}) satisfy the second equation 
(\ref{32c}). We find that this is true if the following constraints on
the constants of integration are satisfied
\begin{equation}
\label{39}
-4C_1^{(1)}C_2^{(1)}n^2 =\alpha^2 \sigma_0 \quad (C_2^{(2)})^2 =
\alpha^2 \sigma_0
\quad \left( (C_1^{(3)})^2 +C_2^{(3)})^2 \right) n^2 =\alpha^2 \sigma_0.
\end{equation}

\end{document}